# Smart Contract Federated Identity Management without Third Party Authentication Services

Peter Mell[1], Jim Dray[2] and James Shook[3]

**Abstract:** Federated identity management enables users to access multiple systems using a single login credential. However, to achieve this a complex privacy compromising authentication has to occur between the user, relying party (RP) (e.g., a business), and a credential service provider (CSP) that performs the authentication. In this work, we use a smart contract on a blockchain to enable an architecture where authentication no longer involves the CSP. Authentication is performed solely through user to RP communications (eliminating fees and enhancing privacy). No third party needs to be contacted, not even the smart contract. No public key infrastructure (PKI) needs to be maintained. And no revocation lists need to be checked. In contrast to competing smart contract approaches, ours is hierarchically managed (like a PKI) enabling better validation of attribute providers and making it more useful for large entities to provide identity services for their constituents (e.g., a government) while still enabling users to maintain a level of self-sovereignty.

**Keywords:** federated identity management; authentication; blockchain; smart contract

## 1 Introduction

Federated identity management (FIM) enables users to access multiple systems using a single login credential. In industry implementations (e.g., with Amazon, Google, and Facebook authentication[4]), multiple entities collaborate such that one entity in the collaboration can authenticate users for other entities; it requires complex interactions to enable a user to perform a business interaction with some 'relying party' (RP) (e.g., a business) and have the authentication performed by a 'credential service provider' (CSP) (the entity performing the authorizations) [TA18]. It may involve redirecting a user from an RP to a CSP and then back to the RP post-authentication with the CSP communicating with both the user and RP. CSPs likely will charge for this service while being able to violate the privacy of users by seeing with which RPs they interact. Complicating matters further, FIM often supports the transferring of user attributes (e.g., age) to an RP to support a business interaction.

---

[1] National Institute of Standards and Technology, Computer Security Division, 100 Bureau Drive Gaithersburg, MD 20899 U.S.A. peter.mell@nist.gov

[2] National Institute of Standards and Technology, Computer Security Division, 100 Bureau Drive Gaithersburg, MD 20899 U.S.A. james.dray@nist.gov

[3] National Institute of Standards and Technology, Computer Security Division, 100 Bureau Drive Gaithersburg, MD 20899 U.S.A. james.shook@nist.gov

[4] Any mention of commercial products is for information only; it does not imply recommendation or endorsement.

In this work, we provide an identity management system (IDMS) that provides FIM such that a user can authenticate and transfer attributes to an RP without the involvement of a CSP (thereby heightening privacy and reducing costs). We accomplish this through leveraging a smart contract running on a blockchain[5]. User to RP interactions do not need to transact with the smart contract, they simply use data from a copy of the blockchain. Thus, there is no need for the user or RP to wait for blockchain blocks to be published or to pay blockchain transaction fees. User to RP communications are extremely fast and free.

Our IDMS is hierarchically managed enabling authorities to manage user accounts and associate attributes with accounts. However, users are granted a degree of self-sovereignty; a user must approve added attributes and can view and delete their data. Privacy is maintained by either adding only hashes of attributes to user records, by only adding data encrypted with the user's public key, or by only adding references to external and secured databases that house user attribute data. We emphasize that user to RP interactions are completely private, something not possible in current systems using a CSP for authentication.

We implemented our IDMS on the Ethereum platform [Eth]. Charges are only incurred when creating and updating user accounts, which is something that is relatively rare compared to a user freely and regularly interacting with RPs. Also, user account update functions are very cheap, all costing less than $0.09 USD (as of September, 2018). We note that other FIM smart contract systems are in development, but ours differs primarily in being a managed approach that still provides a degree of user self-sovereignty. This provides advantages in having authoritative identity attributes for users and having the ability to validate attribute providers.

The rest of the paper is structured as follows. Section 2 describes the overall contract design and section 3 describes the attribute field design. Then section 4 outlines the core functions of the IDMS system: authenticating users and passing attributes. Section 5 provides an example, section 6 discusses our implementation, section 7 explains why we use smart contracts, and section 8 enumerate achieved security properties. Section 9 provides the related work and section 10 our conclusions.

---

[5] See [Yag+18] for an overview of blockchain and smart contract technology.

## 2 IDMS Contract Design

Our IDMS is implemented within a smart contract accessed by five types of entities: the IDMS owner, account managers, attribute managers, users, and RPs (shown in figure 1). The first four issue transactions to the blockchain to manage user accounts (relatively rare events). Users and RPs use public blockchain data to authenticate a user and pass attributes (the more common events). Both the managers and users have IDMS accounts. Manager data is publicly readable while user data is kept private using hashes and encryption.

**Smart Contract**: The smart contract is modeled as being immutable; once deployed, it is not owned and is its own entity. Alternately, it may be coded for the IDMS owner to update it with participant agreement (e.g., a voting mechanism) or after a notification period (allowing participants time to withdraw from the IDMS if they disapprove of the changes).

**IDMS Owner**: The IDMS owner is limited by the contract to authorize and deauthorize managers. For authorization, an entity creates a blockchain account, gives their public key to the owner, and the owner directs the contract to create an IDMS manager account for that public key. For deauthorization, the account record is marked as invalid. For each created manager, the owner specifies one or more descriptor fields. This should follow a standard nomenclature to enable automated evaluation of these fields by other entities (e.g., by RPs).

**Account managers**: Account managers authorize user accounts in an analogous manner as the IDMS owner does for managers. User records are pseudonymous, they contain no identifying information. An account manager can only perform deauthorization on accounts they created. If a user's private key is lost or stolen, the account manager may authorize a new account for the user using a new public key generated by the user and deauthorize the old account. The IDMS owner can require the account managers to perform identity proofing at some level, confirming that users are whom they claim to be. The contract can require a subset of the collected attributes to be posted to the user account. We refer to such attributes as 'identity attributes'; they can be updated at any time by the account manager.

**Attribute managers**: Attribute managers add attributes to users' accounts. However, users must first grant them permission. They may revoke any attributes previously added.

**Users**: Users may unilaterally delete non-identity attributes (to avoid them changing their identity). They may also delete their IDMS account completely. As mentioned previously, they must authorize any attribute manager to add attributes to their account.

**RPs**: RPs keep a local copy of the contract state, extracted from the blockchain, and execute contract 'view' functions on that copy to enable reading the contract data. They do not have accounts on the contract or transact with the contract.

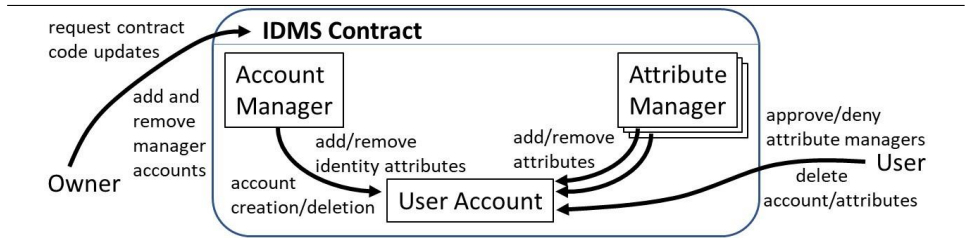

Fig. 1: IDMS Contract Design Relative to a Single User

## 3 IDMS Attribute Field Design

An important design element is the attribute field. Each field has a hash of a user attribute (put there by the applicable account manager or an attribute manager). If the actual attribute data is included to allow for easy user retrieval (which is not necessary), it is encrypted with a secret key that is then encrypted with the user's public key to preserve user privacy. It is expensive to store data on a blockchain; if the data is large (e.g. video or image files), an off-blockchain location of the data may be posted to the attribute field. This might be used, for example, with images of physical credentials such as driver's licenses, visas, social security cards, and passports. Note that the source of each attribute field is public to allow RPs to check the authority behind each user provided attribute.

| Field Name | Field Description |
|---|---|
| ManagerPublicKey | Public key of manager that posted the attribute |
| Identity | Boolean to indicate if this is an identity attribute |
| EncryptedSecretKey | Secret key encrypted with the user's public key |
| Descriptor | Encrypted description attribute data |
| Data | Encrypted attribute data |
| Location | Location for downloading data |
| Hash | Hash of the unencrypted descriptor and data |

Tab. 1: Contents of an Attribute Field

To accomplish this, we use the attribute field structure shown in table 1. The 'ManagerPublicKey' field is the public key of the manager that posted the attribute to the user's account. This key can enable anyone to look up the manager in the IDMS using the publicly available blockchain data. Manager accounts contain only unencrypted attributes so that anyone can verify who posted an attribute. Note that only the contract owner can authorize a manager and populate its data fields, thus the unencrypted attributes within a manager's account are considered authoritative. The 'Identity' field is a boolean indicating whether or not an attribute is an identity attribute. The 'EncryptedSecretKey' is the secret key that was used to encrypt the attribute descriptor and data fields. The 'Descriptor' field is an encrypted field that explains what the attribute data field contains[6]. The optional 'Data' field contains encrypted attribute data (these must be appended with a nonce prior to encryption to prevent guessing attacks when the attribute space is limited). The optional 'Location' field identifies a public location where the encrypted attribute data is available. The 'Hash' field is a hash of the unencrypted Data field appended with the unencrypted Descriptor field. This enables an RP to verify that a user is providing them the correct data and descriptor fields for a particular source. Note that if neither the Data or Location fields are provided, the user must maintain copies of the data for which the relevant hashes are posted.

## 4 IDMS Core Functions

In this section we will describe the core functions for our conceptual IDMS system: 1) authentication of users and 2) secure transmission of user attributes. A key design feature is that the user and RP can achieve this without any interaction with a third party (they don't even need to transact with the smart contract). However, the user needs access to their attribute descriptors and data. These could be maintained by the user, downloaded from the blockchain (if stored in encrypted form in the user's record), or downloaded and decrypted from the location specified in the location field of the user's record. The user will also need to maintain their private key. This could be done in a hardware dongle to promote security and portability between devices, but could also be copied to multiple devices if desired.

The RP will need access to a copy of the blockchain on which the contract is being executed
(which is publicly available through the blockchain peer-to-peer network). They need only store the small portion relevant to the contract data. This must be a version recent enough as to have a hash of the attributes that the user will provide to the RP. Note that the RP does not need a blockchain account and the user will not need to transact with their blockchain account for these core functions (they do so only to maintain their contract user record).

---

[6] Implementations of this should standardize on a set of descriptors and a format for the data field to promote automated processing of the attribute data.

### 4.1 IDMS Authentication

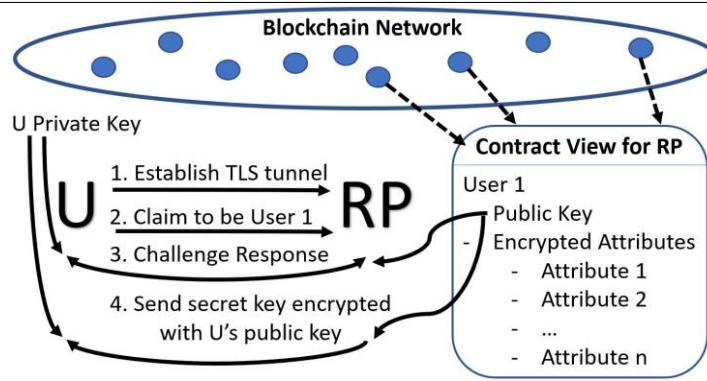

Fig. 2: Example User to RP Authentication Function

Our first core function enables U to authenticate to some $RP_1$ given that $RP_1$ can access U's public key from the IDMS data on the public blockchain. This could be done through many approaches; here we present a method using Transport Layer Security (TLS). Our approach is similar to using TLS with client-side certificates, except that in our scenario no such client-side certificate exists. We achieve this by creating a TLS session, but within that session adding an additional challenge response mechanism followed by $RP_1$ generating a final symmetric key used for a second encrypted tunnel within the original TLS tunnel.

With additional engineering, this tunnel within a tunnel approach could be replaced with the second 'challenge response' tunnel replacing the first TLS tunnel.

More specifically for our example approach, U establishes a TLS tunnel with $RP_1$. U then sends a message to $RP_1$ claiming to own account 'User 1' in the IDMS. $RP_1$ then accesses the IDMS account 'User 1' using its local copy of the blockchain and retrieves the posted public key. $RP_1$ sends a random challenge to U encrypted with the public key posted on the IDMS account. U decrypts this with his private key and sends the result to RP. If the correct value was returned by U, then U has proved ownership of account 'User 1'. Next, $RP_1$ encrypts a symmetric key with U's public key to use for the second encrypted tunnel and sends it to U. U obtains the symmetric key by decrypting with his private key. At this point both U and $RP_1$ have mutually authenticated and have established an encrypted tunnel. This process is shown in figure 2.

Note that in TLS, U produces the symmetric key used for the encrypted tunnel. However, in our secondary tunnel it is necessary that $RP_1$ produce the symmetric key and encrypt it with U's public key to avoid a man-in-the-middle attack. We must prevent $RP_1$ from being able to masquerade as U while accessing some $RP_2$ (because $RP_1$ could answer $RP_2$'s challenge using a response obtained by issuing the same challenge to U).

### 4.2 IDMS Attribute Transfer

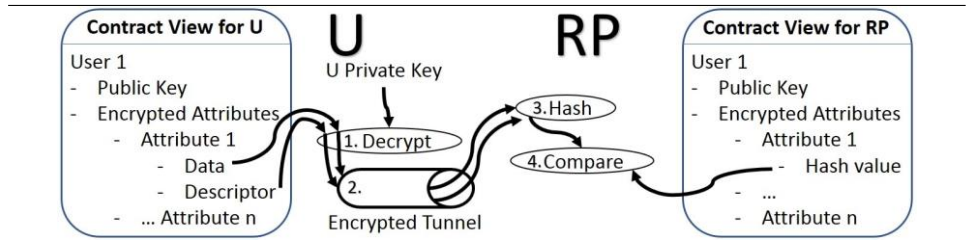

Fig. 3: User to RP Attribute Transfer Function

Our second core function enables U to send attributes to RP (e.g., personal information necessary to complete some interaction). U obtains a decrypted copy of an attribute descriptor and data (from a local store, from an encrypted version stored in the user's IDMS record, or from a server whose location is specified in the user's IDMS record). U sends the descriptor and data to RP. RP hashes a concatenation of the data and descriptor and then verifies that the result matches a hash on the user's IDMS record. The RP can then use the
'ManagerPublicKey' field in the matching attribute record to evaluate the attribute source.

The manager accounts have unencrypted descriptor fields populated by the owner to enable an RP to automatically evaluate the authority of a manager account (e.g., that the manager issuing a drivers license really is the correct government agency). By the owner populating these public manager descriptor fields with a standard nomenclature, automated evaluation by RPs of a manager's authority can be enabled.

## 5 Example Use Case

A government deploys an instance of our IDMS contract to a blockchain and is the owner. The owner authorizes account manager entities to perform identity proofing and add users. This is likely organizations already performing related activities, such as banks and local governments. A user Bob goes to his bank to have an account created in the IDMS. After providing the necessary documentation, he is granted an account. The owner also authorizes a university as an attribute manager with the descriptor fields 'university' and 'University of Corellia'. The former is a standardized descriptor to enable automated processing while the latter provides the name of the specific university (note that how to create ontologies of descriptors is out of scope of this work). Bob then requests that the University of Corellia post his degree to his IDMS account. Bob must first prove to the university using the core IDMS functions that 1) he owns the account and 2) that the account is for his identity by passing them identity attributes. Bob then transacts with the IDMS contract to give the university permission to post attributes to his account. The university gives Bob a digital image of his degree and also posts an attribute on Bob's IDMS account with a hash of the digital image and a location field indicating where Bob can login and download the image off of university servers (in case Bob loses the originally provided digital image). The university posts a second attribute indicating his grade point average (GPA). Since this is a small data field, it is encrypted along with a standard sized nonce and placed inside the attribute field. Bob can download this anytime off of the blockchain and use his private key to decrypt it. Bob then applies for a job with Ally, who wants proof that Bob graduated with a minimum GPA. Bob uses the core IDMS functions to prove that 1) he owns the IDMS account and 2) that the account contains the attributes necessary to convince Ally that Bob received a degree and graduated with a sufficient GPA. When Ally receives and verifies the attributes sent by Bob, she then checks the descriptor fields associated with the attributes. She verifies that the attributes were provided from a university using the first descriptor field and she reads off the specific university using the second descriptor field.

## 6 Implementation Details and Empirical Study

We implemented our IDMS using a smart contract running on the Ethereum platform [Eth] and created apps to interact with the smart contract. The contract implements all of the functionality described in section 2 and it contains methods to support the core functions described in section 4. Note that we left for future work the implementation of the off blockchain U to RP interactions.

We tested all contract interactions described in sections 2 and 4. There were two types of interactions: transactions and views. Transactions are function calls that change the state of the contract; they thus must be submitted to the miners so that the changes can be stored on the blockchain. Views are function calls that look like transactions except that they do not alter the state of the contract; they thus can be executed locally by a node that has a copy of the blockchain. This makes their use free and fast. Table 2 lists the implemented functions.

| Function | Type | Permitted Role | Gas | Ether | USD |
|---|---|---|---|---|---|
| Add Manager | Transaction | Contract Owner | 66632 | 2.0E-4 | $0.03 |
| Delete Manager | Transaction | Contract Owner | 17677 | 5.3E-5 | $0.01 |
| Add User Account | Transaction | Account Manager | 94562 | 2.8E-4 | $0.05 |
| Delete User Account | Transaction | Account Manager | 65020 | 2.0E-4 | $0.03 |
| Add Attribute | Transaction | Managers / Users | 182045 | 5.5E-4 | $0.09 |
| Delete Attribute | Transaction | Managers / Users | 33017 | 9.9E-5 | $0.02 |
| Permit Attribute Manager | Transaction | Users | 45151 | 1.4E-4 | $0.02 |
| Deny Attribute Manager | Transaction | Users | 15283 | 4.6E-5 | $0.01 |
| Compare Hash | View | Public | 0 | 0 | $0 |
| View Attribute | View | Public | 0 | 0 | $0 |
| View Public Key | View | Public | 0 | 0 | $0 |

Tab. 2: IDMS Contract Functions and Costs ($219.01 USD/Ether as of September 27, 2018)

Note that the view functions are used by users and RPs for their interactions. The transaction functions are only used to set up the IDMS data structures. Thus, normal operation of our IDMS is extremely fast and does not cost anything. Creating user accounts and updating them with attributes costs a modest amount of funds (e.g., less than $1 USD), but such activities are relatively rare compared to users interacting with RPs.

## 7 Reasons to use a Smart Contract

Use of the smart contract promotes trust in the system while providing a convenient vehicle for data distribution and update of a distributed and resilient data store. The smart contract code is publicly viewable and immutable, thus all participants know how it will operate and all entities are constrained to their roles. In particular, the owner is limited to just creation and deletion of manager roles; no access to user accounts is provided. The blockchain peer-to-peer network makes it convenient to distribute the IDMS data to participating entities. This also provides transparency and audit-ability for all IDMS transactions. Since the user to RP interactions don't modify the blockchain, this transparency doesn't cause a problem with user privacy. Lastly, the smart contract approach enables one to deploy an IDMS without the need to build and maintain any infrastructure.

## 8 Security Properties

We now summarize the security and privacy properties needed for our model and then explain how each security property is fulfilled by our IDMS and then discuss a residual weakness. The specific security properties are as follows:

1. User attribute data is encrypted such that only the user can decrypt it.
2. Users can securely share their attribute data with other parties.
3. Users can unilaterally remove their attributes.
4. Users can unilaterally remove their account.
5. Users can have multiple accounts in order to hide their association with certain attribute managers.
6. Account managers can only remove accounts that they created. Owners and attribute managers may not remove accounts.
7. Account managers can only modify the identity attributes for accounts they created.
8. Attribute managers may only place attributes if explicitly permitted by the relevant user.
9. Owners may only add and remove account/attribute managers and update the IDMS contract code.
10. IDMS contract code may only be updated by the owner following due process laid out in the contract (which is publicly available to all users of the contract).

11. Relying parties can trust account managers to perform identity proofing that binds real world entities to user accounts at a stated level of assurance.

These security properties are provided primarily by the contract itself. Except under conditions documented within the contract, the code is immutable. The code is also public so that users can verify that these properties will be held. The contract directly enforces security properties 3, 4, 6, 7, 8, 9, and 10. Key to this enforcement is for the smart contract to authenticate the party requesting a change. This is handled by the smart contract system, leveraging the accounts on the blockchain. Thus, our approach does not have to implement that part of the trust model.

Property 1 is enacted by the account and attribute managers when they place attributes on a user account. There is nothing in the contract to prevent the posting of unencrypted attributes, but there is no motive for a manager to do so and there could be repercussions (e.g., the owner could remove the manager from the IDMS).

Property 2 is enabled since our IDMS architecture provides a way for a user and RP to directly authenticate and pass attributes. All they need is to use a standard encrypted connection within which to execute our protocol.

Property 5 can be provided by a user's account manager. It is trivial to create additional accounts on blockchain systems, thus the user can do so easily. The account manager then simply creates an IDMS account with the public key associated with each of the user's accounts. Based on our empirical work, there may be a modest cost to create each account (e.g., $0.05 USD). Also, we note that users are not required to pass RPs their identity attributes, enabling them to pass other attributes without revealing their identity. This can enable transactions to authenticate that a person has some attribute while staying anonymous. An example might be an online forum where only members of a certain organization can post messages but where the poster's identity is to remain anonymous.

Property 11 is achieved through the contract owner auditing the account managers to ensure that users are identity proofed at the required or advertised level of assurance. If account managers are non-compliant then the contract owner can revoke their accounts.

Despite these security protections, we note an important limitation. An account managers could use their knowledge of a user's identity attributes to create a clone identity for someone else. This is analogous to a government duplicating someone's passport but including a different picture to enable someone to act as someone else. To our knowledge this problem exists in the related schemes (discussed next) whenever attribute managers act maliciously.

## 9 Related Work

Many organizations are investigating using blockchain technology for identity management. Our approach is unique in providing a managed hierarchical approach with user selfsovereignty that can authoritatively validate attribute providers (or claim providers).

**uPort**: uPort is an 'open identity system for the decentralized web' [uPo18]. uPort users create and manage self-sovereign identities by creating Ethereum accounts linked to a self-sovereign wallet. Being unmanaged and fully self-sovereign, there is no entity identity proofing of user accounts [Lun+17]. Our approach differs in that it provides a managed solution that still provides a level of self-sovereignty. This managed aspect can enforce validation on the claim providers not possible in completely unmanaged systems.

**SCPKI**: The paper entitled 'SCPKI: A Smart Contract-Based PKI and Identity System' [AlB17] addresses the issue of rogue certificates issued by Certificate Authorities in traditional public key infrastructures. It proposes an alternative PKI approach that uses smart contracts to build a decentralized web-of-trust. The web-of-trust model is adopted from the Pretty Good Privacy (PGP) system [Gar95]. SCPKI supports self-sovereign identity by defining a smart contract that allows users to add, sign, and revoke attributes. Users can sign other user's attributes, gradually building a web-of-trust where users vouch for each others' identity attributes. As with uPort, our approach differs in that it provides a managed model that can provides additional assurances on claims.

**Ethereum Improvement Proposal 725**: Ethereum Improvement Proposal 725 [Vog17] (EIP-725) defines a smart contracts based identity management framework where each identity account is a separate smart contract. It supports self-attested claims and third party attestation. EIP-725 is augmented by EIP-735 [Vog], which specifies standard functions for managing claims and is supported by the ERC-725 Alliance [ERC]. An online ERC-725 DApp demonstration is available [0RI]. Our approach has similar capabilities but does not require every user and issuer of claims to have their own smart contract; ours is also a hierarchical managed model.

**Sovrin**: Sovrin is 'a protocol and token for self-sovereign identity and decentralized trust' [Sov]. Its goal is to replace the need for PKIs and to create a Domain Name System (DNS) type system for looking up public keys to be used for identity management purposes through building a custom blockchain system. It is a permissioned based cryptocurrency with no consensus protocol, thus it has centralized ownership of the tokens. The managing Sovrin foundation must approve all nodes managing the blocks but is appealing for community involvement in running nodes. The token is a cryptocurrency so that value can be exchanged along with supporting identity transactions. Our approach differs in that it doesn't require its own blockchain or cryptocurrency and can be executed on top of any smart contract system.

**Decentralized Identity Foundation**: The Decentralized Identity Foundation (DIF)is a large partnership with the stated goal of building an open source decentralized identity ecosystem [Fou18]. The primary focus is on high level framework, organizational issues, and standards. DIF plans to develop a broad, standards based ecosystem that supports a range of different implementations.

**Other Related Work**: There are many other FIM related blockchain projects that cannot be referenced here due to space limitations. For the majority of them, the design details are unavailable or are in constant flux due to the nascent nature of this market.

## 10 Conclusions

We have demonstrated that it is possible to design a FIM system that enables direct user to RP authentication and attribute transfer without the involvement of a third party. We implemented this using a smart contract and identified the advantages of taking such an approach. We note that user to RP interactions do not require transactions with the contract, making them fast, free, and private.

Our approach provides strong user self-sovereignty so that only the user can view and share their attribute data. However it is a managed system, intentionally not fully self-sovereignty as with the cited related work to prevent users from unilaterally changing their own identity and to provide greater validation of attribute providers. Our limits on self-sovereignty also enable the IDMS to provide authoritative and consistent data about users and participating organizations. Our approach is thus suitable for a large organization to provide identity management services to its constituents (e.g., a government). Once established by a large entity, other organizations may leverage the IDMS to provide attributes to their users and gain the ability to identify and authorize users (but only with user permission). If the owner of the contract opens up the system to many account managers and attribute managers, this will create a powerful identity management ecosystem (as opposed to being a service only for a particular purpose).